\documentclass[reqno]{amsart}
\usepackage[pagewise]{lineno}\linenumbers
\usepackage{graphicx}
\usepackage{caption}
\usepackage[labelformat=simple,labelfont={}]{subcaption}
\usepackage{bm}
\usepackage{float}
\usepackage{amsmath}
\usepackage{mathtools}
\usepackage{eqparbox}
\usepackage{color}
\usepackage{multirow}
\usepackage{placeins}
\usepackage{amsmath}
\usepackage{nameref,hyperref}
\usepackage{xurl} 
\usepackage{epstopdf}

\theoremstyle{definition}

\theoremstyle{remark}

\numberwithin{equation}{section}

\title{Interplay between Foraging Choices and Population Growth Dynamics}

\author[J. Calvo-Monge]{Jimmy Calvo-Monge}
\address{Centro de Investigación en Matemática Pura y Aplicada, Universidad de Costa Rica, San Jose, Costa Rica}
\email{jimmy.calvo@ucr.ac.cr}

\author[B. Espinoza]{Baltazar Espinoza}
\address{Biocomplexity Institute, University of Virginia, Virginia, USA}
\email{baltazar.espinoza@virginia.edu}

\author[F. Sanchez]{Fabio Sanchez}
\address{Centro de Investigación en Matemática Pura y Aplicada - Escuela de Matemática, Universidad de Costa Rica, San Jose, Costa Rica}
\email{fabio.sanchez@ucr.ac.cr}

\begin{document}
\nolinenumbers




\keywords{Population dynamics, mathematical modeling, population growth, logistic model, adaptive behavior}

\begin{abstract}
In this study, we couple a population dynamics model with a model for optimal foraging to study the interdependence between individual-level cost-benefits and population-scale dynamics. Specifically, we study the logistic growth model, which provides insights into population dynamics under resource limitations. Unlike exponential growth, the logistic model incorporates the concept of carrying capacity, thus offering a more realistic depiction of biological populations as they near environmental limits. We aim to study the impact of individual-level incentives driving behavioral responses in a dynamic environment. Specifically, explore the coupled dynamics between population density and individuals' foraging times. Our results yield insights into the effects of population size on individuals' optimal foraging efforts, which impacts the population's size.
\end{abstract}
\maketitle

\allowdisplaybreaks 
\section{Introduction}

The logistic population growth model is an essential framework in ecology, providing a realistic depiction of how populations grow under resource constraints. Unlike exponential growth, which presumes limitless resources, the logistic model incorporates the concept of carrying capacity--the maximum population size an environment can sustain.

Initially developed by Pierre François Verhulst in 1838~\cite{Verhulst38}, the model was formulated to better understand the dynamics of population growth as it encounters environmental limits~\cite{krebs2001}. The logistic population growth equation is mathematically represented as:
    \begin{equation}\label{eq:classicLogistic}
        \frac{dP}{dt} = rP \left(1 - \frac{P}{K}\right)
    \end{equation}
where \(P\) denotes the population size at time \(t\), \(r\) denotes the intrinsic growth rate, and \(K\) stands for the environment-specific carrying capacity. The factor \(\left(1 - \frac{P}{K}\right)\) modulates the growth rate and encapsulates the reduction in growth as the population approaches \(K\)~\cite{brauer2012}.

The foundational work of renowned ecologists and biologists such as Simon A. Levin, David Tilman, and Mark E. J. Newman underscores the logistic growth model's critical importance in population dynamics. Simon Levin's ecological and evolutionary dynamics research has highlighted how logistic models can effectively capture the complex interplay between individual behaviors and population-level outcomes \cite{levin1992problem}. David Tilman's studies on plant competition and biodiversity have utilized logistic growth models to reveal the underlying mechanisms that drive species coexistence and ecosystem stability \cite{tilman1982resource}. Mark Newman's contributions to statistical physics and complex systems have extended the applicability of logistic growth models to broader contexts, demonstrating their utility in understanding the growth and spread of populations in interconnected networks \cite{newman2003structure}. The enduring relevance of these pioneering works lies in their ability to provide a robust mathematical framework for exploring how populations grow, compete, and stabilize over time. For biologists and other researchers, the insights gained from these studies offer valuable tools for predicting and managing population dynamics in diverse ecological and evolutionary scenarios. Moving forward, leveraging the logistic growth model with contemporary data and advanced computational techniques will enable researchers to address pressing biological and ecological challenges with greater precision and predictive power.

The logistic population growth is a pivotal concept not just in ecology, where it emerged backed by substantial empirical evidence \cite{Feller40}, but also in fields such as economics \cite{KWASNICKI201350, Smirnov2017InSO} and epidemiology \cite{ZHAO2023107,SAHA2021}, where it aids in understanding and predicting the dynamics of systems constrained by finite resources. The S-shaped (or sigmoid) logistic curve starts with exponential growth, progresses through a deceleration phase, and stabilizes at the carrying capacity, providing a comprehensive view of population dynamics over time \cite{turchin2003}. Given the importance of the logistic function, its mathematical properties and the characteristics of growth processes have been deeply explored~\cite{MEYER1999247,KUCHARAVY2015280}.

Understanding the logistic growth model is fundamental for effective decision-making in various sectors, including conservation, urban planning, and resource management. Its ability to forecast changes in systems limited by essential resources makes it a valuable tool for these sectors. Given the great importance and applicability of the logistic growth model, the discussion around its implications, assumptions, and limitations is very rich, and it encompasses interpretations and opinions around various topics such as density-dependent selection and evolutionary ecology, see for example \cite{mallet2012struggle}. One of the crucial observations on the logistic growth general model is the relationship between reproduction fitness and population density. Under the logistic model formulation, the intrinsic growth rate is the main parameter that describes this relationship as it relates to the ability of its agents to reproduce and accelerate population growth. In classical logistic models, such as~\eqref{eq:classicLogistic}, this biological indicator is kept constant, thus assuming an invariant health capability to reproduce, which is unaffected by the status of the environment in which the agents are placed. There are various mathematical generalizations of the classical model \eqref{eq:classicLogistic}, which aim to provide more flexibility through different growth parameters~\cite{TSOULARIS200221, Buis1991-BUIOTG}; usually, these formulations provide a fixed formula to compute the growth rate using fixed parameters and the population density as inputs. Another example of the linkage between population density and reproductive fitness is the well-known Allee effect \cite{Dennis89, COURCHAMP1999405}, which assumes an inverse density relation as the population crosses a minimum population size threshold.

The previously mentioned modeling approaches formulate density-dependent growth by incorporating attributes observed at the population scale. These frameworks overlook the effects of individual-level reproductive and behavioral traits on shaping population-level growth.
%
%
In reality, individuals' survival and reproductive abilities are expected to evolve dynamically depending on the population size, which is affected by self-decisions. Individuals' characteristics, such as energy consumption, 
physical and health indicators (body size and temperature), modulate population growth~\cite{Savage04}.
The classical logistic population growth formulation fails to account for individual-level decision factors that can impact aggregated population dynamics.
Reconciling individual-level incentives with population-level dynamics requires a framework that explicitly incorporates the intertwined dynamics between population growth and individual decision-making processes modulating reproductive fitness.
%

In their pioneering work, Clark and Mangel developed a series of models to study the optimal behavioral decision of agents seeking to increase their fitness in fragmented environments~\cite {mangel1988dynamic,clark2000dynamic}. In their modeling approach, a foraging individual is confronted with a prey badge decision process, where each batch offers an associated energy gain, an energy cost, and a probability of effectively finding food. By employing a finite planning horizon approach, the decision process leads to an optimal batch selection process fed by all of these constraints.

We present an adaptive logistic growth model largely inspired by this approach. Under our formulation, a representative agent of the population is confronted every day with the following behavioral decision problem: how much effort should the agent invest in trying to find food for the day? This problem is mainly framed around two factors: $(i)$ the energy gained if food is found for each time effort available, and $(ii)$ the probability of finding food for each time effort. By incorporating these factors, the proposed adaptive approach yields an optimal decision on the daily effort the agent should make to find food. The decision-making process returns the agent's remaining energy level, affecting their reproductive fitness and ultimately modulating the intrinsic growth rate.
In other words, the population growth rate results from aggregated individual behavioral processes considering the cost-benefit trade-offs of seeking resources to survive and reproduce. The decision-making process comes down to finding the optimal amount of time and energy consumption an agent has to invest in looking for resources to survive, henceforth called ``foraging time''.
The two factors affecting the continuous decision-making process are not static; they depend on the growth of the population. In this case, the probability of finding food per foraging time effort might intuitively decrease with larger population sizes as resource acquisition becomes competitive. This is very common in adaptive formulations where the local decisions and the global scale intertwine: the agent makes a decision constantly, which is informed on the overall system status, but the system continuously progresses using the updated decision parameters obtained at the individual level.

In this case, the effort to obtain food is modeled through the foraging time. Hence, this behavioral framework drives the decision agent to modify its foraging time selection per day to adapt to the population changes and their direct impact on the availability of resources and the overall probability of finding food in the environment. An advantage of this methodology is that the agent's sensitivity to change their foraging time selection can be incorporated into the mathematical formulation. Consequently, this approach allows for a more granular description of the population members. Compared with the classical model, the intrinsic growth of the population is constantly updated depending on the population size. However, the update is non-centralized, resulting from a purely individualized optimization process.

In this paper, We present the mathematical formulation of this approach in section \ref{sec:methods}. In section \ref{sec:results}, we present the main numerical differences that this approach entails with respect to the classical approach. We will see that not only can we obtain different population-level behaviors, but we can also obtain a more detailed depiction of the effort made by the constituent agents to support population growth. In section \ref{sec:discussion}, we discuss the main implications of this approach, which we believe are of great importance in highlighting the benefits of incorporating individual optimizations that play into the progress of global biological processes.

\section{Methods}\label{sec:methods}

The behavioral approach modulates the intrinsic population growth rate $r$ through time. This is done by considering the daily energy level of the foraging individual, $e_t$. This way, the global population growth becomes linked with the reproductive fitness of its constituent individuals. In this model, the intrinsic reproduction rate is modified proportionally to the individual daily level, that is, $r_t = r_c \cdot e_t$, where $r_c$ is a constant that relates the individual's fitness with their capacity for procreate and, thus, increase the population size.

The representative agent must select the total foraging time (time looking for food) spent daily. After this decision, the cost of foraging time and possible energy gain will determine the agent's final daily energy level, $e_t$. Hence, this method involves a daily iteration of the logistic growth equation with a renewed intrinsic growth ($r$) parameter. 

The mechanism to select the optimal daily foraging time depends on several factors, including the current population size (in principle, a larger population means more competition to obtain food resources). We base this method on the adaptive approach first presented in \cite{fenichel2011adaptive}, utilizing
employing a dynamic optimization algorithm to obtain the desired optimal decision.

We use a Markov decision process formulation to perform the decision process at the optimal daily foraging time. In this process, we consider a set of states $\mathcal{S}$ with two values $\{FF, NFF\}$ (Finding Food, Not Finding Food). We let the foraging agent have a utility $u^h(f)$ of engaging in $f$ units of foraging time at day $t$ for each state $h \in \mathcal{S}$. The individual can choose foraging time units in an acceptable range $f \in \mathcal{F} = [f_{\min}, f_{\max}]$. Following the traditional adaptive approach, we model this utility function to be of the shape $u^h(f) = \alpha^h(b^hf-f^2)^\nu$ where $b=2f_{\text{opt}}$, where $f_{\text{opt}}$ is the foraging time selection that produces the maximum immediate utility assuming food is indeed found at this foraging time selection. We may use a normalization parameter $\alpha^h$ to keep the utility values in a given range. 

\begin{figure}
\includegraphics[scale=0.45]{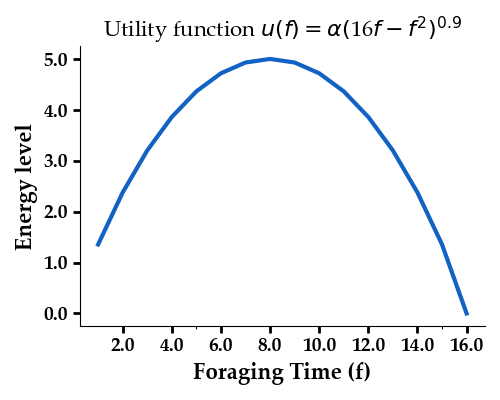} 
\caption{Example of energy gain per foraging time selection. In this case, the agent has an optimal foraging time $f_{\text{opt}}=8$ and a sensitivity parameter of $\nu=0.9$. The parameter $\alpha$ is computed so that this function attains a maximum energy value of $u=5$.}
\label{fig:UtilityFunctionExample}
\end{figure}

In this case, the utility function can be interpreted as the net energy gain for each amount of foraging time units. The curved concave shape of the utility function -see Figure (\ref{fig:UtilityFunctionExample})- tells us that increasing foraging time investment can lead to higher energy gains up to a point where the effort of taking more time units looking for food implies consuming more energy. The parameter $\nu$ controls the slope of this function. Higher $\nu$ creates utility functions with higher slopes, commonly interpreted as an individual with a lower sensitivity or propensity to change their optimal foraging time decision (as changing it might incur higher energy loss).

We assume each foraging time selection $f$ has a probability of finding food $P^{FF}(f,P(t))$, which also depends on the size of the population at time $t$ (that is, the value of $P(t)$). Intuitively, this probability should be an increasing function in $f$ (as more foraging time could lead to a higher possibility of finding food) but should also have lower values when the population size increases. This is because more individuals introduce competition for resources, making it more difficult to obtain food at a given foraging time selection over time. There is not a single way to model these requirements mathematically. We provide an example of a mathematical formulation for this function in the appendix, where we explore the crucial effect that the values of the $P^{FF}(f,P(t))$ function have on the overall model results.

Any Markov decision process relies heavily on the use of expected utilities, which inform the agent of the net utility to gain throughout the projection horizon. For each state, we have the corresponding value functions or expected utilities $V^{FF}_t$ and $V^{NFF}_t$ computed for $t$ in a time horizon $t \in [1, \tau]$. The expected utility of finding food is obtained mathematically through the Bellman equation.

\small{
\begin{align}\label{eqn:bellman-find-food}
    V^{FF}_t = \max_{f \in \mathcal{F}} \{ 
    u^{FF}(f) + \delta \left[ P^{FF}(f) V^{FF}_{t+1} + (1-P^{FF}(f))\cdot V^{NFF}_{t+1}\right]
    \} \quad \text{for} \quad t=1,2,\cdots,\tau-1.
\end{align}}

This equation formalizes the search for the optimal foraging selection by considering the immediate utility $u^{FF}(f)$ and discounting the future utility by a discount factor of $\delta$. The expected utility of not finding food is computed more simply by considering the Bellman equation

\begin{align}\label{eqn:bellman-not-find-food}
    V^{NFF}_t = \max_{f \in \mathcal{F}} \{ 
    u^{NFF}(f) + \delta V^{NFF}_{t+1}
    \} \quad \text{for} \quad t=1,2,\cdots,\tau-1.
\end{align}

For the case of the $NFF$ state, the Bellman equation will yield the optimal selection $f_{\text{opt}}$ in each iteration of the planning horizon. This is because the value of $V^{NFF}_{t+1}$ is independent of $f$, so the optimization is carried on a translation of the utility function $u^{NFF}(f)$. Equation (\ref{eqn:bellman-not-find-food}) is added to ensure the theoretical completeness of the framework. However, the real optimization process happens considering the immediate utility and the expected utility of finding food (the $FF$ state), which is encapsulated in equation (\ref{eqn:bellman-find-food}).

Equations (\ref{eqn:bellman-find-food}) and (\ref{eqn:bellman-not-find-food}) are solved with a backward induction methodology, starting with $V_{\tau}^{FF}=u(f_{\text{opt}})$ (the value of the utility function at the immediate optimal foraging time selection) and $V_{\tau}^{NFF}=0$. The optimal foraging time selection is computed as the final iteration result $f^*= \text{argmax}V_t^{FF}$ at $t=1$, and the resulting daily energy gain corresponds to the value of the immediate utility at the optimal foraging time selected, that is $e=u^{FF}(f^*)$. 

\section{Results}\label{sec:results}

\begin{figure}[H]
    \includegraphics[scale=0.4]{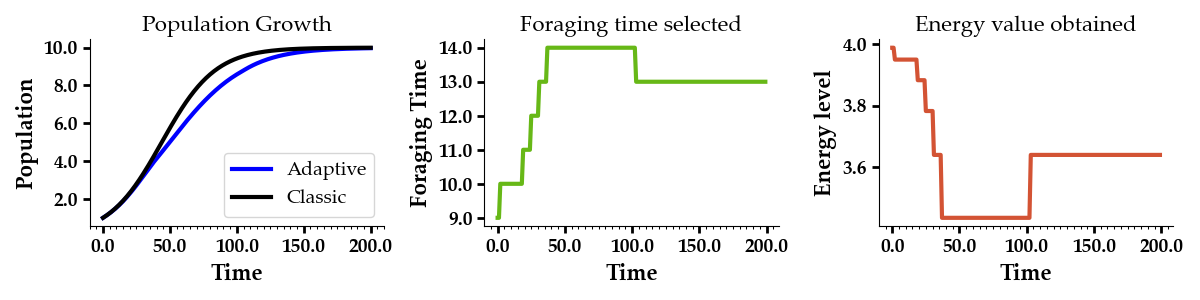}
    \caption{Population growth, optimal daily foraging time selected and net daily energy gain for an adaptive logistic growth process with parameters $\nu=0.4, \tau=7, \delta=0.98, b=16, k=10$ and $r_c=0.01$. The utility function attains a maximum value of $u=5$ at $f_{\text{opt}}=8$.}
    \label{fig:PopGrowthAndFrgTime_1}
\end{figure}

\begin{figure}[H]
    \centering
    \includegraphics[scale=0.4]{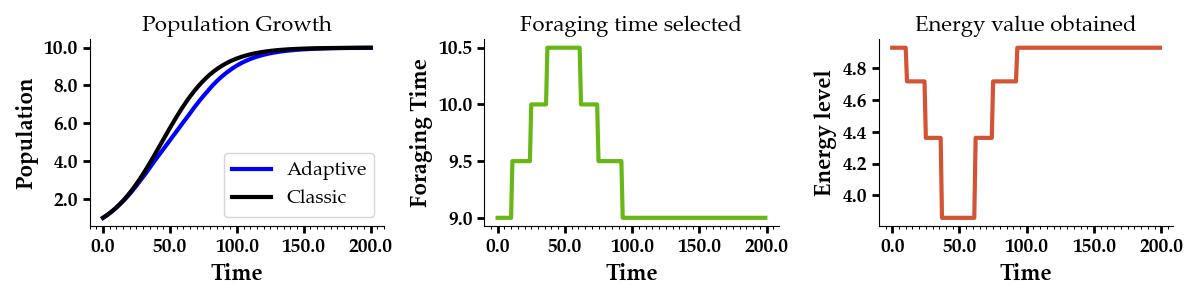}
    \caption{Population growth, optimal daily foraging time selected and net daily energy gain for an adaptive logistic growth process with parameters $\nu=0.8, \tau=7, \delta=0.98, b=16, k=10$ and $r_c=0.01$. Increasing the $\nu$ parameter reduces the sensitivity of the decision agent, thus creating a lower deviation from the optimal immediate foraging time selection.}
    \label{fig:PopGrowthAndFrgTime_2}
\end{figure}

We start by observing the effect on the population dynamics obtained with the adaptive approach and the progress of the optimal foraging time selection and resulting energy levels obtained. Figure (\ref{fig:PopGrowthAndFrgTime_1}) displays the results of a simulation of the adaptive logistic process versus a classical logistic process where $r=r_c \cdot u(f_{\text{opt}})$ in which the agent never deviates from its optimal immediate foraging time decision. In this scenario, the agent has a maximum immediate utility (energy gain) of $u=5$ when selecting $f_{\text{opt}}=5$ units of foraging time. We can see that throughout the adaptive process, the increase in population size compels the decision agent to exert an increasing effort by spending more time looking for food. The energy loss of this effort is evidenced by a shift away from the maximum utility value. When the population limit is attained, the decision becomes stale at a foraging time higher than the immediate optimal, at $f^*=14$ foraging time units. Under this scenario, the high population density demands a larger effort of the agent to continue with acceptable energy levels. By contrast, Figure (\ref{fig:PopGrowthAndFrgTime_2}) presents a similar simulation using a higher $\nu$ parameter, which reduces the sensitivity of the decision agent. This results in a decision process in which agents are less likely to deviate from the optimal foraging time $f_{\text{opt}}$ and are less strained at the limited population density.

Figures \eqref{fig:PopGrowthAndFrgTime_1}
and \eqref{fig:PopGrowthAndFrgTime_2} display one of the most elucidating conceptual results that can be obtained via the adaptive framework: \textit{the ability to visualize two process scales: the individual efforts and their impact in the global population system}. In the first scenario, we have a population with high sensitivity to change their foraging times. This means these agents must engage in more extreme efforts to preserve sufficient energy levels. In contrast, we face a more resilient population in the second scenario. This framework makes These fitness descriptions available through the utility function shape, which ultimately describes the a priori trade-off between foraging time effort and energy gain.

As a result of these sensitivity differences, the foraging times selections and the obtained energy differ throughout population growth. The first, most sensitive population is forced to increase their effort in finding food because of their sensitivity to changes in conditions. Thus, they experience a reduction in their energy levels. The second population, the more resilient to change, can afford to return to their optimal decision and retain their energy levels.

Both scenarios show us a way to quantify the strain or burden placed on the population to survive, a feature that is virtually hidden if one only examines the evolution of global population growth. Even though both populations achieve the same total capacity in similar timeframes, one gets there with significantly more energy drain than the other. 

\begin{figure}[H]
    \centering
    \includegraphics[width=0.9\linewidth]{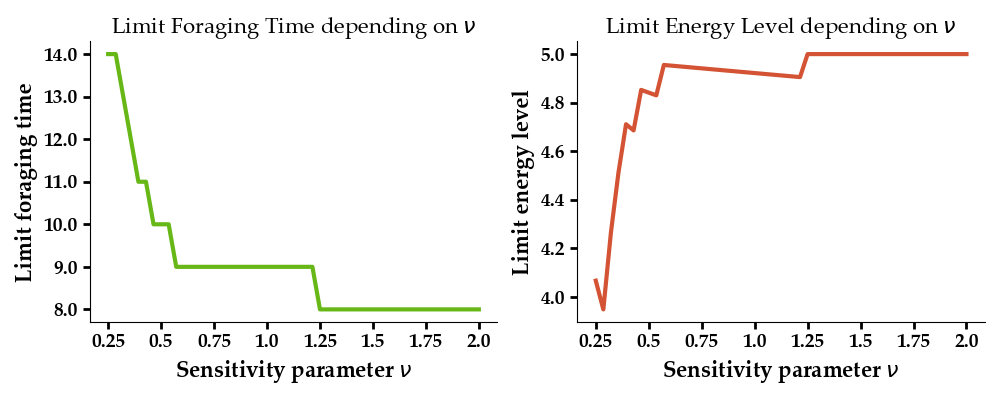}
    \caption{Limit Foraging Times and Energies depending on the sensitivity parameter $\nu$. Using $\tau=7, \delta=0.98, b=16, k=10$ and $r_c=0.01$, and a utility function that has a maximum value of $u=5$ in all cases, attained at $f_{\text{opt}}=8$.}
    \label{fig:LimitsDependingOnNu}
\end{figure}

To further illustrate this point, Figure \eqref{fig:LimitsDependingOnNu} shows the limit foraging times and energy levels present at the population when it reaches its total capacity, depending on the sensitivity parameter $\nu$. Increasing $\nu$ -making agents less sensitive to change- reduces the effort needed and increases their final energy gains.

\begin{figure}[H]
    \centering
    \begin{tabular}{cc}
         \includegraphics[width=0.45\linewidth]{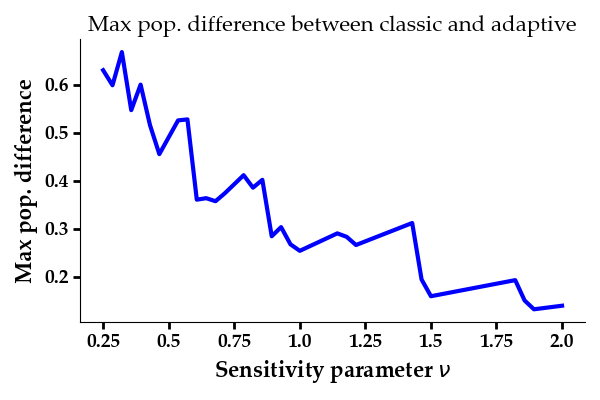} & 
         \includegraphics[width=0.5\linewidth]{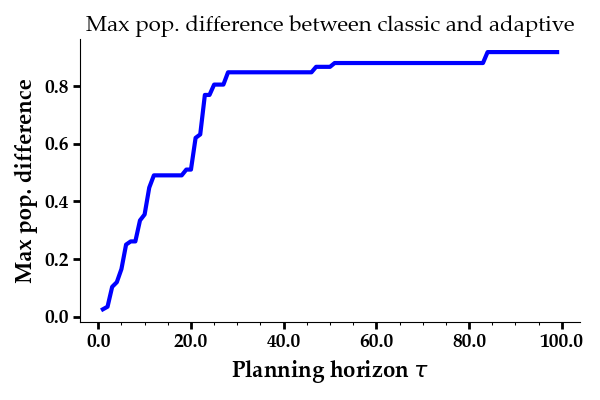}
    \end{tabular}
    \caption{Maximum population difference between classical logistic with $r_t=r_c*u(f_{\text{opt}})$ constant and the adaptive logistic formulation. On the left, the difference depends on the sensitivity parameter $\nu$; on the right, the planning horizon $\tau$ varies. Using $\delta=0.98, b=16, k=10, r_c=0.01$, a value of $\tau=14$ for the left panel and a value of $\nu=0.5$ for the right.}
\label{fig:DifferenceWithClassicDependingOnNuTau}
\end{figure}

Besides the further description of the energy state of the agents that constitute the population, the adaptive framework also yields population curves that attenuate the growth of the overall logistic curve with respect to the classic case in which individuals make no selection. This is because by changing the foraging time selection, the agent deviates from their a priori optimal energy gain, and they lose energy because of the pressure that increasing food competition exerts over their decision rationale. We can see this difference, depending on adaptive parameters $\nu$ and $\tau$ in Figure \eqref{fig:DifferenceWithClassicDependingOnNuTau}, where, like before, more sensitive populations yield higher differences with respect to the classical scenario. 

\textbf{Remark}. The numerical results become unstable for much lower sensitivity parameter values $\nu$, and the adaptive approach does not provide a substantial difference from the classical approach. In these cases, the utility function is almost flat (very low slope), implying no major energy difference across foraging time options. We add the values of $\nu$, which give us meaningful qualities of the adaptive approach with respect to the classical formulation. In extreme cases where the population is not sensitive (higher values of $\nu$), the difference with the classical logistic becomes null: there is no major incentive to adapt. The adaptive framework provides a window through which individuals susceptible to adapting to circumstances can influence the global population system.

\section{Discussion}\label{sec:discussion}

The results of our study highlight the significant interplay between individual-level incentives and population-scale dynamics within the framework of logistic growth. By integrating a behavioral model for optimal foraging with the logistic growth model, we provide a comprehensive view of how individual decisions can affect and be affected by population dynamics.

Our findings underscore the importance of considering individual behavior in ecological models. Traditional logistic growth models, which assume a constant intrinsic growth rate, overlook the dynamic nature of individual decision-making and its impact on population growth. By allowing the intrinsic growth rate to be modulated by individual energy levels, our adaptive model offers a more realistic depiction of population dynamics. This approach recognizes that resource availability, competition, and individual health influence reproductive fitness and population growth rates.

The adaptive framework illustrates that individuals must invest more effort in foraging to obtain the same resources as population size increases, leading to higher energy expenditures. This increased foraging effort reduces the net energy gain and thus affects the intrinsic growth rate of the population. Our results show that the optimal foraging time increases with population size, reflecting the heightened competition for limited resources. This finding aligns with the principle that higher population densities result in reduced per-capita resource availability, needing behavioral adaptations for survival.

Additionally, our study offers insights into the energy dynamics within populations. We observed that populations with higher sensitivity to environmental changes (i.e., lower \(\nu\) values) exhibited increased foraging times and energy expenditures. Conversely, populations with lower sensitivity (higher \(\nu\) values) could maintain optimal energy levels with less deviation in foraging times. This distinction underscores the adaptive framework's ability to capture individuals' nuanced trade-offs to balance energy intake and expenditure in response to population density and resource competition.

Furthermore, the adaptive framework quantitatively measures the strain or burden placed on individuals within a population. We can assess the effort required for survival and reproduction under different population densities by examining the foraging time and energy levels. This level of granularity in modeling offers a deeper understanding of the individual contributions to population growth and overall health.

The versatility of the adaptive framework utilized in this method is another significant outcome of our study. The adaptive dynamic programming approach for individual optimal decision-making has proven to be a valuable tool for implementing decision-making processes at the individual level while integrating with global population systems. Traditionally, this framework has been explored in the context of infectious disease modeling \cite{espinoza2021adaptive,espinoza2022heterogeneous,fenichel2011adaptive, fenichel2013economic,morin2013sir,CALVOMONGE2023769}. In this article, we extend its application to enhance logistic growth processes. However, the fundamental concept can be applied to any global biological process influenced by parameters determined through individual-level optimizations. We encourage the community to investigate the benefits of the adaptive framework in bridging agent-level optimizations with the progress of global biological systems.

In conclusion, integrating individual-level behavioral models with population-scale dynamics presents a more comprehensive and realistic approach to understanding population growth. By accounting for individual decision-making processes and their impact on reproductive fitness, our adaptive logistic growth model provides valuable insights into the complex interplay between individual actions and population trends. This approach opens new avenues for ecology, resource management, and conservation research, offering a robust framework for predicting and managing population dynamics in the face of environmental challenges.


\section*{Acknowledgements}
The authors, J.C.M. and F.S., are grateful for the support of the Research Center in Pure and Applied Mathematics and the Department of Mathematics at the University of Costa Rica. This work was partially supported by the National Science Foundation (NSF) through Expeditions in Computing Grant CCF-1918656 and NSF through Incorporating Human Behavior in Epidemiological Models (IHBEM) grant DMS-2327710.
Supporting code for the article can be found in \href{https://github.com/JimmyCalvoMonge/adaptiveLogistic}{this repository}.

\newpage
\appendix

\section{Shape of the probability of finding food function}

\begin{figure}[H]
    \centering
    \includegraphics[scale=0.6]{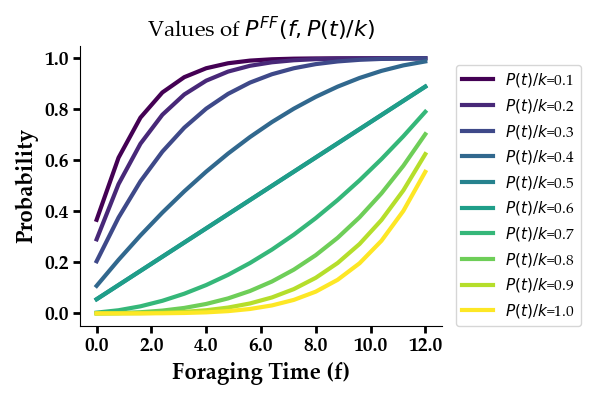}
    \caption{Values of the function $P^{FF}$ for each foraging time selection, as the population size $P(t)$ increases (We use the proportion of population size per carrying capacity $P(t)/K$ as the parameter to modify $P^{FF}$). Overall probabilities increase with higher foraging times, however the magnitude and increase speed become lower as the population grows. These curves were obtained using the Beta C.D.F. approach.}
    \label{fig:ProbFindingFoodPerPop}
\end{figure}

As mentioned in the main article, the probability of finding food should increase with the foraging time selection, but it ought to have lower values as the population density increases (this is an intuitive interpretation of the process that population densities play in the competition for food). This mathematical component comes with greater flexibility in our model proposal, and in this section, we present a proposal to formulate this function. Figure (\ref{fig:ProbFindingFoodPerPop}) shows the proposed behavior of the $P^{FF}(f,P(t))$ function.

These function shapes can be obtained through several mathematical artifacts. In our case, we used the shape provided by the cumulative distribution functions of Beta distributions. Specifically, the C.D.F. from the Beta distribution can be obtained as 
\begin{align*}
    F(x;\alpha,\beta) = \frac{B(x;\alpha, \beta)}{B(1;\alpha, \beta)}, \quad \text{where} \quad
    B(x;\alpha, \beta) = \int_0^x t^{\alpha-1}(1-t)^{\beta-1}dt
\end{align*}
is the incomplete Beta function, \cite{deGroot12}. The $\alpha, \beta$ parameters can control the steepness of the curve, therefore we can use the value of $P(t)$ to modify them. The function is then translated to have domain $\mathcal{F}=[f_{\min}, f_{\max}]$. Another approach to creating the shape of the $P^{FF}(f,P(t))$ function can be the use of sigmoid functions.

\section{Utility function shapes}

\begin{figure}
\includegraphics[scale=0.4]{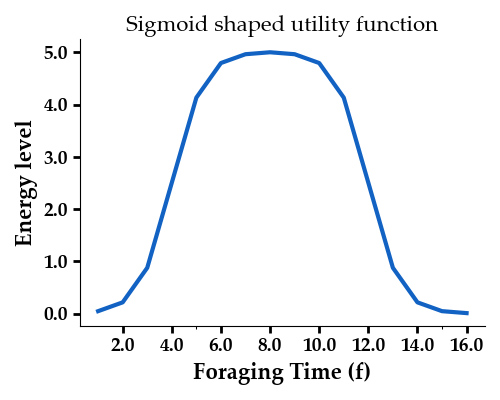}
    \caption{Utility function with sensitivity parameter $\theta=1.55$ obtained as a reflection of two sigmoid functions across the $f=f_{\text{opt}}$ axis.}
\label{fig:UtilityFunctionExampleSigmoid}
\end{figure}

The utility shape is also a mathematical property of the adaptive approach, which can be modified with versatility. The standard quadratic function presented in the main article and first proposed in \cite{fenichel2011adaptive} can be replaced with similar shapes. For example, a sigmoid function of the shape 
\begin{align}\label{eq:SigmoidEquation}
    s(f) = \alpha/(1+e^{\theta(f_{\text{opt}}-f)})
\end{align}
used to create a utility function just as in Figure \eqref{fig:UtilityFunctionExampleSigmoid}. In this case, a higher sigmoid parameter, $\theta$, yields a more sensitive to change population (unlike the case in the quadratic formulation where higher $\nu$ parameters are interpreted as less sensitive populations). 

\begin{figure}[H]
    \centering
    \includegraphics[scale=0.4]{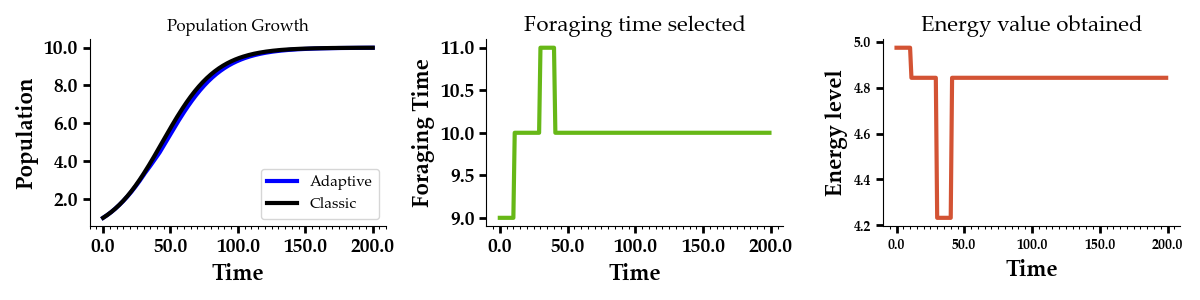}
    \caption{Population growth, optimal daily foraging time selected, and net daily energy gain for an adaptive logistic growth process with parameters $\tau=7, \delta=0.98, b=16, k=10$ and $r_c=0.01$. The utility function attains a maximum value of $u=5$ at $f_{\text{opt}}=8$ and uses a sigmoid function with a sensitivity parameter of $\theta=1.7$.}
    \label{fig:PopGrowthAndFrgTime_Sigmoid_2}
\end{figure}

\begin{figure}[H]
    \centering
\includegraphics[scale=0.4]{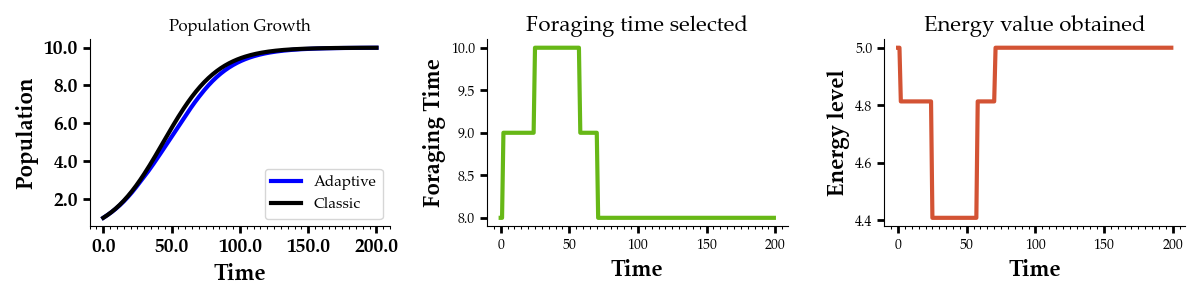}
    \caption{Population growth, optimal daily foraging time selected, and net daily energy gain for an adaptive logistic growth process with parameters $\tau=7, \delta=0.98, b=16, k=10$ and $r_c=0.01$. The utility function attains a maximum value of $u=5$ at $f_{\text{opt}}=8$ and uses a sigmoid function with a sensitivity parameter of $\theta=0.9$.}
\label{fig:PopGrowthAndFrgTime_Sigmoid_1}
\end{figure}

We conduct simulations similar to the main article but employ this new utility function shape. We note that the adaptation is perhaps less evident or severe with this utility shape. Figure \eqref{fig:PopGrowthAndFrgTime_Sigmoid_1} shows a sensitive population forced to keep a higher energy effort, and Figure \eqref{fig:PopGrowthAndFrgTime_Sigmoid_2} shows a population with a lower sensitivity, which can resume optimal energy levels in the limit. This is a similar situation to the one discussed in our main results.

Although the adaptive method's formulation follows the standard procedures of Markov Decision Process modeling, this example shows that the model is very sensitive to hyperparameters and function shapes. This allows for greater model versatility to capture different ecological scenarios and population attributes; however, it increases the cation needed for an accurate modeling methodology. The example in the next section also speaks to this sensitivity, specifically to how the modeler chooses to formulate the probability of finding food for each foraging time option.

\section{Sensitivity to the probability of finding food}

\begin{figure}[H]
    \centering
    \includegraphics[scale=0.6]{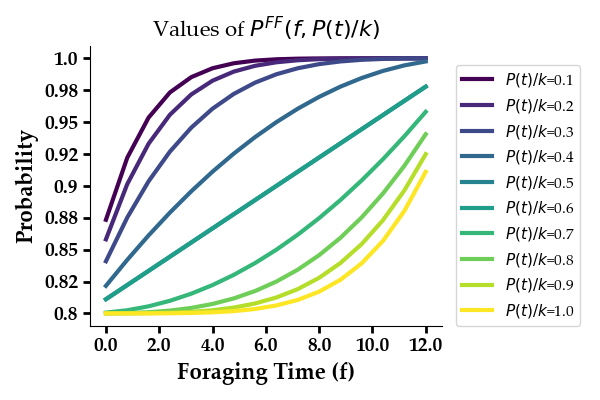}
    \caption{Values of the function $P^{FF}$ for each foraging time selection, as the population size $P(t)$ increases, using a the same approach as in Figure \eqref{fig:ProbFindingFoodPerPop} however adding a minimum value of $P=0.8$.}
    \label{fig:ProbFindingFoodPerPop2}
\end{figure}

As we have formulated, the probability of finding food for each foraging time selection depends on the changing status of the population dynamics. In Figure \eqref{fig:ProbFindingFoodPerPop}, we offer a heuristic approach to define such probabilities a priori. The definition of these probability functions is a very important factor that influences the outcome of the adaptive process. For example, if we increase the general probability values to have a minimum value of $P=0.8$, as in Figure \eqref{fig:ProbFindingFoodPerPop2}. 

\begin{figure}[H]
    \centering
\includegraphics[scale=0.4]{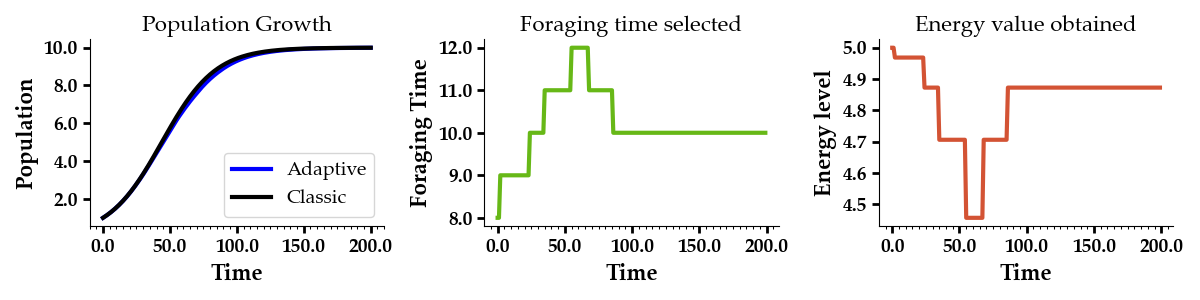}
    \caption{Population growth, optimal daily foraging time selected and net daily energy gain for an adaptive logistic growth process with parameters $\nu=0.4, \tau=7, \delta=0.98, b=16, k=10$ and $r_c=0.01$. In this case, we use the probability of finding food $P^{FF}(f, P(t))$ given by Figure \eqref{fig:ProbFindingFoodPerPop2}, which has higher overall probabilities. Compare this to the corresponding simulation in the main article, which uses the same parameters. Higher probabilities of finding food incur in less energy demands from decision agents.}
\label{fig:PopGrowthAndFrgTime_3}
\end{figure}

A process using the same parameters of Figure (2) of the main article but using this new $P^{FF}$ function gives a very different result, as seen in Figure \eqref{fig:PopGrowthAndFrgTime_3}, where the energy strain placed on the foraging agents is less severe (precisely because of the increase in the probability of finding food for each foraging selection). The circumstances described by Figure \eqref{fig:ProbFindingFoodPerPop2} reflect a population with more advantageous conditions than one with probabilities given by Figure \eqref{fig:ProbFindingFoodPerPop}. These variations could result from environmental or communal attributes, which make it more/less probable to find food at a given foraging time.

\vskip2pc
\bibliographystyle{RS}
\bibliography{references}

\end{document}